\documentclass[letter,twocolumn]{aa}
\usepackage[colorlinks=true,urlcolor=blue,citecolor=blue]{hyperref}
\usepackage{epstopdf}
   
\begin{document}

\title{Near-infrared detection and characterization of the exoplanet HD\,95086\,b with the Gemini Planet Imager\thanks{based on public data taken at the GPI commissioning.}}

\author{R. Galicher\inst{\ref{inst1}}, J. Rameau\inst{\ref{inst2}}, M. Bonnefoy\inst{\ref{inst2}}, J.-L. Baudino\inst{\ref{inst1}}, T. Currie\inst{\ref{inst3}}, A. Boccaletti\inst{\ref{inst1}}, G. Chauvin\inst{\ref{inst2}}, A.-M. Lagrange\inst{\ref{inst2}}, C. Marois\inst{\ref{inst4}}}
\institute{LESIA, CNRS, Observatoire de Paris, Univ. Paris Diderot, UPMC, 5 place Jules Janssen, 92190 Meudon, France\email{raphael.galicher@obspm.fr} \label{inst1}\and UJF-Grenoble 1 / CNRS-INSU, Institut de Plan\'etologie et d'Astrophysique de Grenoble (IPAG) UMR 5274, Grenoble, F-38041, France \label{inst2}\and Department of Astronomy and Astrophysics, Univ. of Toronto, 50 St. George St., Toronto, ON, M5S 1A1, Canada\label{inst3}\and National Research Council of Canada Herzberg, 5071 West Saanich Road, Victoria, BC, V9E 2E7, Canada \label{inst4}}

\titlerunning{HD\,95086 GPI follow-up}
\authorrunning{Galicher et al.}

\abstract{
HD 95086 is an intermediate-mass debris-disk-bearing star.VLT/NaCo $3.8~\mu m$ observations revealed it hosts a $5\pm2\,\mathrm{M}_{Jup}$ companion (HD\,95086\,b) at $\simeq 56$\,AU.   Follow-up observations at 1.66 and 2.18 $\mu m$ yielded a null detection, suggesting extremely red colors for the planet and the need for deeper direct-imaging data.  In this Letter, we report H- ($1.7~\mu m$) and $\mathrm{K}_1$- ($2.05~\mu m$) band detections of HD\,95086 b from Gemini Planet Imager (GPI) commissioning observations taken by the GPI team.  The planet position in both spectral channels is consistent with the NaCo measurements and we confirm it to be comoving. Our photometry yields colors of H-L\,'= $3.6\pm 1.0$ mag and K$_1$-L\,'=$2.4\pm 0.7$ mag, consistent with previously reported 5-$\sigma$ upper limits in H and Ks. The photometry of HD\,95086\,b best matches that of 2M\,1207\,b and HR\,8799\,cde.  Comparing its spectral energy distribution with the BT-SETTL and LESIA planet atmospheric models yields T$_{\mathrm{eff}}\sim$600-1500\,K and log\,g$\sim$2.1-4.5.   Hot-start evolutionary models yield M=$5\pm2$\,M$_{Jup}$. Warm-start models reproduce the combined absolute fluxes of the object for M=4-14\,M$_{Jup}$ for a wide range of plausible initial conditions (S$_{init}$=8-13\,k$_{B}$/baryon).  The color-magnitude diagram location of HD\,95086\,b and its estimated T$_{\mathrm{eff}}$ and log\,g suggest that the planet is a peculiar L-T transition object with an enhanced amount of photospheric dust.  
}  


\date{ }

\maketitle

\section{Introduction}
HD 95086 b is a directly imaged  planet ($5\pm2$\,M$_\mathrm{J}$, $a_{proj}$ = 55.7 $\pm$ 2.5\, $AU$)  discovered by \citet{rameau13a} in L$^\prime$ (3.8 $\mu m$) with VLT/NaCo \citep{lenzen03,rousset03}  orbiting the young A8 star HD\,95086 (M$\sim$ 1.6 M$_{\odot}$), a member of the Lower Centaurus Crux subgroup \citep[$17\pm4$\,Myr,][]{pecaut12,meshkat13}.  Additional L$^\prime$ images taken later in 2013  confirmed that the object is comoving with its star  \citep{rameau13b}.

NaCo Ks ($2.18~\mu$m) and NICI \citep{chun08} H-band ($1.65~\mu$m) observations failed to reveal the planet \citep{rameau13a,meshkat13}.  However,   $5\,\sigma$ lower limits of Ks-L$^\prime=1.2\pm0.5$ mag and H-L$^\prime=3.1\pm0.5$\,mag suggest that the planet may have extremely red colors, similar to the young planets HR 8799 bcde and 2M 1207 b \citep{Chauvin04,Marois08,Marois10a}, which have very dusty/cloudy atmospheres \citep{Barman2011,Currie2011}.  Higher contrast near-IR data able to detect HD 95086 b can provide better comparisons with these objects and better constrain its atmosphere.

In this Letter, we present detections of HD\,950\-86\,b with the recently installed Gemini Planet Imager \citep[GPI,][]{macintosh14} on Gemini South from public data as a part of GPI commissioning observations \citep{perrin14}.  The data (acquired and reduced by the GPI team), their analysis, and the detections are presented in \S\,\ref{sec: obs}. In \S\,\ref{sec: photoastro},  we combine GPI H and $\mathrm{K}_1$ photometry with NaCo L$^\prime$ photometry to constrain the physical properties of HD\,95086\,b.

\section{Observations and data reduction}
\label{sec: obs}
The GPI is a new instrument for imaging and characterizing planets around young nearby bright stars,  combining an extreme adaptive optics system, coronagraphs, and an integrated field spectrograph (IFS). The IFS platescale is $14.3\pm0.3~\mathrm{mas}.\mathrm{px}^{-1}$ for a $2.8''$ field-of-view (FOV) and the true North position angle is given within 1\,deg\footnote{http://planetimager.org/\label{foot: gpi}}.
\begin{table*}[htbp]
\centering
\caption{Observing log of HD\,95086 with GPI}
\begin{tabular}{cccccc}
\hline\hline
Date&Filter&Coro mask diam (mas)&DIT(s)$\times$NDITS$\times$Nb$_\lambda$&Nb images&FOV rotation ($^\circ$)\\
\hline
$2013/12/10$&$\mathrm{K}_1$-coro&306&$119.29278\times1\times37$&17&11.7\\
\hline
$2013/12/11$&H-coro&246&$119.29278\times1\times37$&21&15.0\\
\hline
\end{tabular}
\tablefoot{\sl Date, filter, occulting mask diameter, exposure, numbers of coadds, of spectral channels, of images, and FOV rotation.}
\label{tab: obslist0}
\end{table*}

The HD\,95086 spectral data were obtained at H ($1.5-1.8\,\mu$m, R=$44-49$) and $\mathrm{K}_1$ ($1.9-2.19\,\mu$m, R=$62-70$) in 2013 December using apodized Lyot coronagraphs (Tab.\,\ref{tab: obslist0}) and angular differential imaging \citep[ADI,][]{marois06a}. Conditions were good:  0.43\arcsec and 0.6\arcsec DIMM seeing, air masses of $1.32$ and $1.34$, and coherence times of $19$\,ms and $17$\,ms, respectively.   The GPI commissioning team used their pipeline for bad-pixel removal, destriping, non-linearity and persistence corrections, flat-fielding, wavelength calibration, and converting the data into spectral data cubes. We used the data cubes relying on the GPI pipeline quality.  The data are made of $21$ and $17$ spectral cubes at H and $\mathrm{K}_1$ bands, respectively, consisting of $37$ spectral channels each.

\begin{figure}[htbp]
\centering
\includegraphics[width=.85\columnwidth]{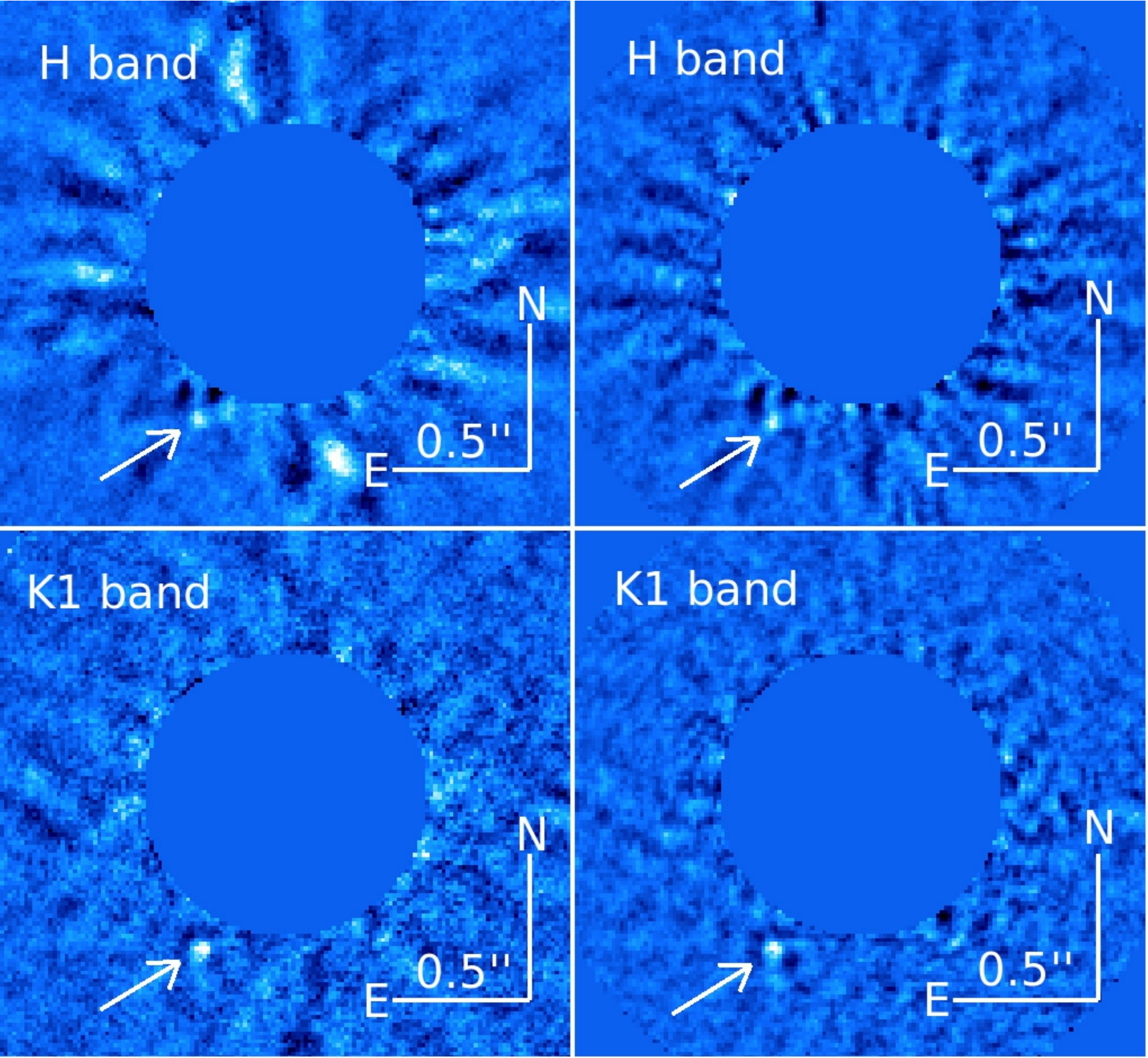}
\caption{\sl Final images of the HD\,95086 system at H (top) and $\mathrm{K}_1$ (bottom) bands from two of our pipelines. The planet (arrow) is detected in all images. The bright speckles are masked up to 500\,mas from the central star.}
\label{fig: im}
\end{figure}

To further process the data, we registered each slice of the spectral cubes using the barycenter of the four satellite spots \citep[attenuated replica of the central star PSF induced by a grid placed in a pupil plane,][]{marois06b}. Then, we minimized the speckle noise in each slice using independent pipelines each adopting various methods \citep{marois06a,lafreniere07,lagrange10,boccaletti12,chauvin12,soummer12,currie13,marois14} used for ADI and spectral differential imaging \citep[SSDI,][]{racine99}. Finally, all slices were mean-combined to yield an integrated broad-band image to maximize the signal-to-noise ratio (S/N) of any off-axis source. Binning images in wavelength and suppressing the speckles (ADI), or suppressing the speckles in each spectral channel (ADI/ADI+SSDI) and binning images give similar results, and all our pipelines recover HD\,95086\,b, which is the sole bright spot at the expected separation. Thus, we provide the first detections at H and $\mathrm{K}_1$ bands (Fig.\,\ref{fig: im}) with an S/N of $\sim$ 3-4 and 5-6, respectively. The central bright speckles are masked up to 500\,mas.  These are the first detections of HD 95086 b with an instrument than is not NaCo/VLT. No spectrum can be extracted given the low S/N.

To estimate the planet flux and position, we needed unsaturated GPI PSFs. As GPI cannot acquire off-axis observations of the star, we calibrated photometry and astrometry using the satellite spots, which are expected to have same shape and brightness for a given filter. In the laboratory the spot-to-central-star flux ratios were $2.035\times10^{-4}$ (9.23\,mag) and $2.695\times10^{-4}$ (8.92\,mag) at H and $\mathrm{K}_1$ bands\textsuperscript{\ref{foot: gpi}}. To check these values, we compared H and K photometry of HD\,8049\,B \citep[VLT/NACO-SINFONI, ][]{zurlo13} with our measurements derived from public GPI HD 8049 data. Assuming that the object is not photometrically variable with time and considering the laboratory spot contrasts, GPI and VLT photometry are consistent within $\epsilon_1=0.2$\,mag, which we take as the error on the ratios. From these ratios, we assessed biases induced by our processing by injecting fake point-sources (i.e., unsaturated PSFs) into the data before applying speckle-suppression techniques \citep{lagrange10,Marois10b,chauvin12,galicher12}. We obtained templates of the planet image. Adjusting the flux of the templates, we found the planet photometry and the fitting error $\epsilon_2$, which depends on the detection quality. $\epsilon_2$ is 0.8\,mag and 0.3\,mag at H and $\mathrm{K}_1$. Finally, we estimated the variation $\epsilon_3$ of stellar flux over the sequence with the variation of spot flux. $\epsilon_3$ is $0.2$\,mag and $0.3$\,mag over the H and $\mathrm{K}_1$ sequences including the variations between spots. The resulting photometric error is the quadratic error, which is dominated by the low S/N at H and is a mix of all errors at $\mathrm{K}_1$.

\begin{figure}[htbp]
\centering
\includegraphics[width=8cm]{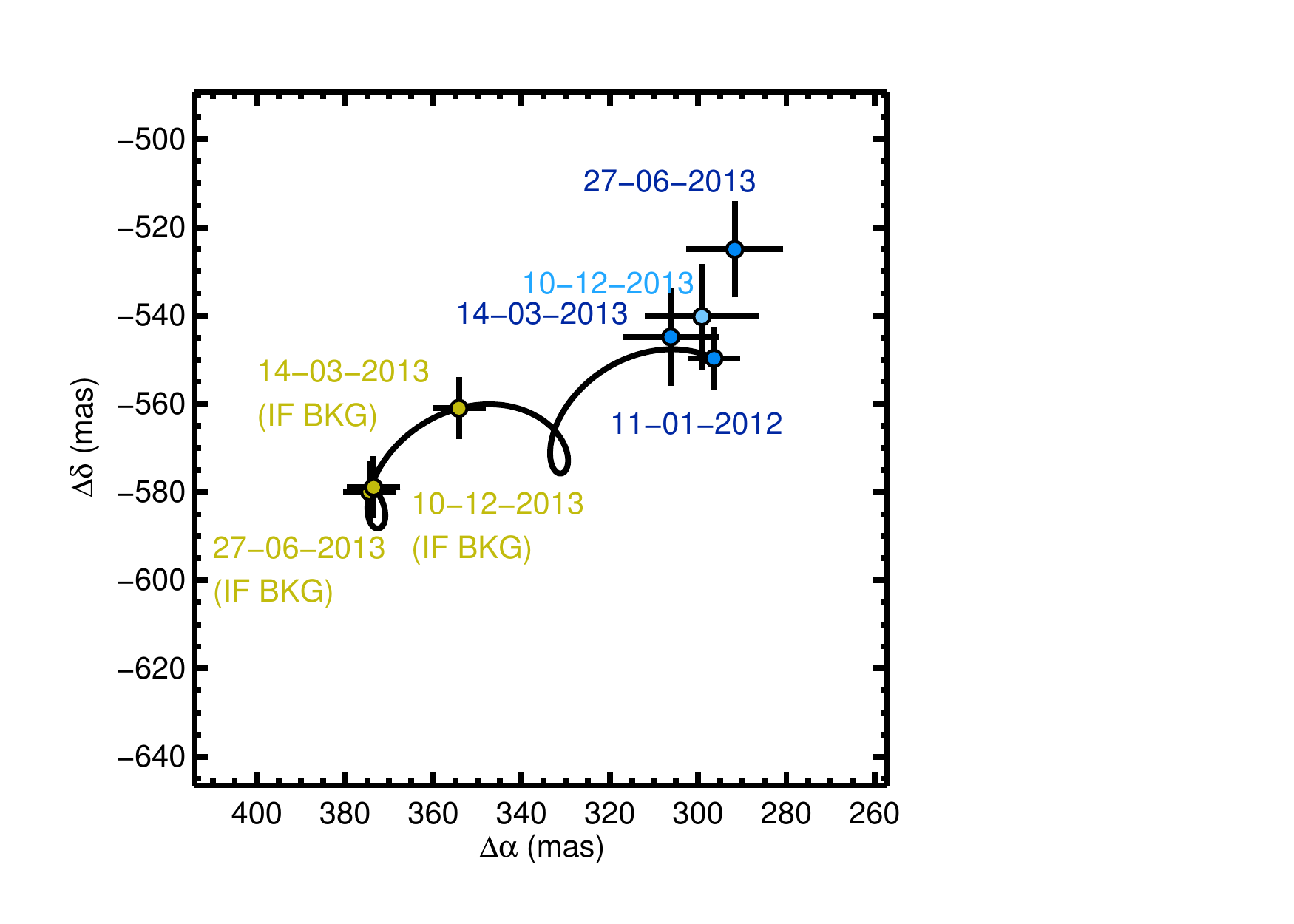}
\caption{\sl HD\,95086\,b positions from its star in RA ($\Delta\alpha$) and DEC ($\Delta\delta$). GPI and NaCo measurements are marked in blue and expected positions of a background object in yellow.}
\label{fig:radec}
\end{figure}

For the astrometric error, we considered uncertainties in the centroiding accuracy of individual slices ($\le0.3$\,pixel), the plate scale  ($0.02$\,pixel), the planet template fit ($0.7$\,pixel at H, $0.5$\,pixel at $\mathrm{K}_1$), and the North position angle (1\,deg). The error is dominated by the low S/N of the detections and the generic GPI calibrations. The current precision is good enough to assess the comoving status of the companion (Fig.\,\ref{fig:radec}). We tried to use the astrometric standard HD\,8049\,B in GPI data to better constrain the North orientation. We did not succeed because of the high orbital motion of HD\,8049\,B and because there is no contemporary observation from other instruments.

Final measurements are presented in Tab.\,\ref{tab: photastro}.  We include revised  2012 NaCo L$^\prime$ photometry obtained by  1) better calibrating the planet signal \citep[as in ][]{currie13} and 2) precisely deriving the L$^\prime$ neutral density (ND) filter throughput (used to flux-calibrate HD 95086) by comparing ND and unsaturated $\beta$ Pic data.
\begin{table} 
\caption{\sl HD\,95086\,b photometry and astrometry at H and $\mathrm{K}_1$ (GPI data) and L$^\prime$ \citep[and revision*]{rameau13a,rameau13b}.}
\label{tab: photastro}
\begin{tabular}{ccccc}
\hline\hline
Date&Filter&Sep (mas)&PA($^{o}$)&$\Delta$m\\
\hline
\tiny 2013/12/11&\tiny H&\tiny$633\pm17$&\tiny$150.6\pm1.7$&\tiny$13.1\pm0.9$\\
\hline
\tiny 2013/12/10&\tiny$\mathrm{K}_1$&\tiny$623\pm15$&\tiny$151.4\pm1.5$&\tiny$12.1\pm0.5$\\
\hline
\tiny 2013/03/24&\tiny L$^\prime$&\tiny$626.1\pm12.8$&\tiny$150.7\pm1.3$&\tiny$9.71\pm0.56$\\
\hline
\tiny 2012/01/11&\tiny L$^\prime$&\tiny$623.9\pm7.4$&\tiny$151.8\pm0.8$&\tiny$9.79\pm0.40$\\
\hline
\tiny 2012/01/11&\tiny L$^\prime$&\tiny$623.9\pm7.4$&\tiny$151.8\pm0.8$&\tiny$9.48\pm0.19*$\\
\hline
\end{tabular}
\end{table}

\section{Characterization}
\label{sec: photoastro}
Absolute magnitudes were derived from the contrast ratios (Tab.\,\ref{tab: photastro}): $M_{H}=15.29\pm0.91$\,mag, $M_{K_1}=14.11\pm0.51$\,mag, and M$_{L^\prime}=11.44\pm0.22$\,mag using the 2MASS and WISE W1 \citep{cutri03,cutri12} photometry of the star\footnote{Correction factors from the GPI/NaCo and 2MASS/WISE photometry, derived from  the spectrum of an A7III star in the \cite{pickles98} library, are negligible.}.

Combining the H band GPI data with revised NaCo L$^\prime$ data, we compared the L$^\prime$/H-L$^\prime$ color-magnitude diagram position of HD\,95086\,b with that of young companions, field dwarfs \citep{Leggett10,Leggett13}, and LYON evolutionary tracks \citep{Chabrier00,Baraffe03} generated for the GPI/NaCo passbands\footnote{http://phoenix.\-ens-lyon.fr/\-simulator/\-index.faces}.
\begin{figure}
\includegraphics[width=.9\columnwidth]{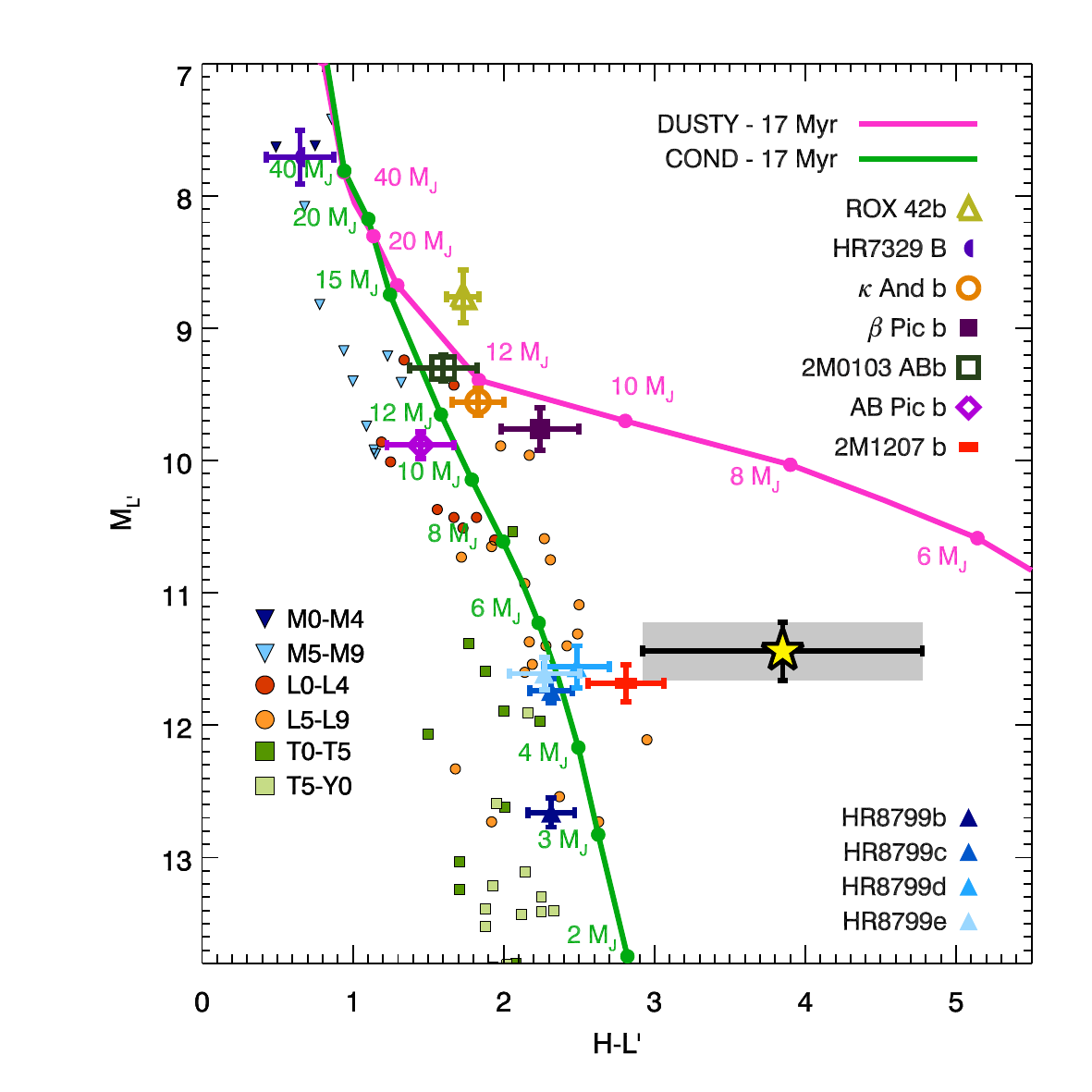}
\caption{\sl Color-magnitude diagram using the new H-band photometry of HD\,95086\,b (yellow star) and data from \citet{Bonnefoy13}, \citet{Bonnefoy14}, and 
\citet{currie14}.}
\label{fig:CMD}
\end{figure}
We converted the GPI measurements into H photometry by applying correction factors derived from published spectra, the filter transmissions, and a spectrum of Vega. HD\,95086\,b lies at the L-T transition in this diagram, similar to other young (8-30\,Myr) planets like HR\,8799\,cde \citep{Marois08, Marois10a} and 2M1207\,b \citep{Chauvin04}. Its red H-L$^\prime$ color compared with the sequence of field dwarf objects \citep{Leggett10,Leggett13} suggests a high content of photospheric dust \citep{Barman2011,Currie2011}, owing to reduced surface gravity \citep[e.g. Fig.\,11 of][]{Marley12}.

We built the 1.5-4.8\,$\mu$m spectral energy distribution (SED) of the planet following \citet{Bonnefoy13} by combining the GPI photometry with the L$^\prime$ one. The normalized SED (at L$^\prime$) is best compatible with the young exoplanets HR\,8799\,bcde and 2M1207\,b, but is redder. Its colors are also $\sim$1 mag redder than those of the benchmark dusty L6.5-L7.5 field dwarf 2MASS J22443167+2043433 \citep{Dahn02,Stephens09}.

We also compared the SED of  HD\,95086\,b with the predictions from grids of synthetic spectra for BT-SETTL \citep{Allard12} and LESIA atmospheric models \citep{Baudino14}.  Each synthetic SED was normalized to that of HD\,95086\,b at L$^\prime$. The BT-SETTL grid covers $\mathrm{400\:K \leq T_{\mathrm{eff}} \leq 3500\:K}$ with 50 to 100\,K increments, $\mathrm{-0.5 \leq log\:g \leq 6.0}$\,dex with 0.5\,dex increments, and M/H=0.0 or +0.5\,dex. The BT-SETTL models that reproduce the photometry of HD\,95086\,b have $\mathrm{600\:K \leq T_{\mathrm{eff}} \leq 1500\:K}$ and  $\mathrm{3.5\:dex \leq log\:g \leq 4.5\:dex}$. The three LESIA grids assume $\mathrm{700\:K \leq T_{\mathrm{eff}} \leq 2100\:K}$, $\mathrm{2.1 \leq log\:g \leq 4.5}$\,dex, and solar abundances: one without clouds and two with clouds of Fe and Mg$_2$SiO$_4$ particles. For each LESIA model, we selected the planet radius that minimizes $\chi ^2$ between the observed and calculated apparent magnitudes. We only kept models with a radius in a realistic range derived from evolution models \citep[0.6 to 2 Jupiter radii,][]{mordasini12}.  All LESIA models that reproduced the HD\,950866\,b photometry have $\mathrm{900\:K \leq T_{\mathrm{eff}} \leq 1500\:K}$ and  $\mathrm{2.1\:dex \leq log\:g \leq 4.5\:dex}$.

The planet mass cannot be derived from the atmosphere models, and evolutionary models are needed. Comparing the planet's L$^\prime$ luminosity with hot-start DUSTY and COND models for an age of $17\pm4$ Myr, we find a planet mass of M=$5\pm2$\,M$_{Jup}$ (Tab.\,\ref{tab:masses}). We did not use the H and $\mathrm{K}_1$ photometries because they are poorly reproduced by the models for an object at the L-T transition (larger uncertainties than for L').
\begin{table} 
\caption{Physical parameters predicted by hot-start evolutionary models for the observed absolute magnitudes.}
\label{tab:masses}
\begin{tabular}{ccc|cc}
\hline\hline
&\multicolumn{2}{c|}{SED}&\multicolumn{2}{c}{L'}\\
Model&BT-SETTL&Lesia&Dusty&Cond\\
\hline
$\mathrm{T_{\mathrm{eff}}\:(K)}$&$1050^{+450}_{-450}$&$1200^{+300}_{-300}$&$916^{+43}_{-44}$&$1108^{+66}_{-65}$\\
$\mathrm{log\:g\:(dex)}$ &$4.0^{+0.5}_{-0.5}$&$3.3^{+1.2}_{-1.2}$&$3.8^{+0.1}_{-0.1}$&$3.9^{+0.1}_{-0.1}$\\
M ($\mathrm{M_{Jup}}$)&--&--&$4.5^{+1}_{-1}$&$5.5^{+1.5}_{-1.5}$\\   
\hline
\end{tabular}
\end{table}
 The models predict T$_{\mathrm{eff}}$ and log\,g, in agreement with those derived from the SED fit.

\begin{figure}
\includegraphics[width=.85\columnwidth]{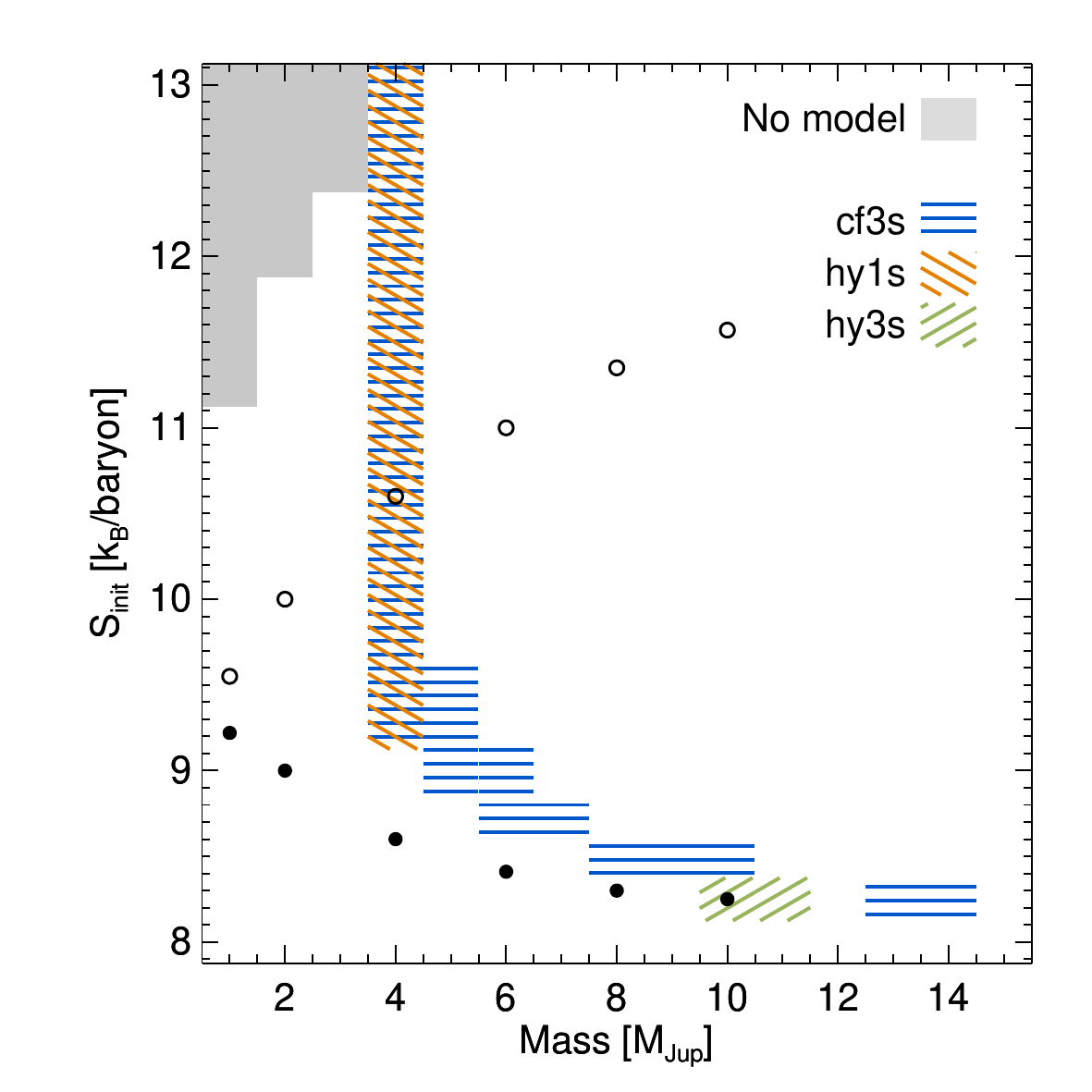}
\caption{\sl Combination of initial entropies ($S_{init}$) and masses (shaded areas) for which the planet 1.6-4.8\,$\mu$m photometries are reproduced by the warm-start models of \cite{SB12} withing 1\,$\sigma$. Three boundary conditions are considered:  with (hy) and without (cf) cloudy atmospheres,  at solar (1s) and 3x solar (3s) metallicity. Initial entropies for the cold-start (filled circles) and hot-start (open circles) models of \cite{marley07} are overlaid.}
\label{fig: WS}
\end{figure}

Alternatively, we used the warm-start models \citep{SB12} to account for possible different  initial conditions for the planet (parameterized by the initial entropy between 8 and 13 k$\mathrm{_{B}}$/baryon). The models assume solar metallicity and atmospheres enriched by a factor of 3 with/without dust clouds as boundary conditions. Synthetic SEDs are generated from predicted spectra of planets\footnote{http://www.as\-tro.prince\-ton.edu/\~bur\-rows/warm\-start/spec\-tra.tar.gz}. For the full range of initial entropies we considered, models assuming masses of $4-14\,\mathrm{M_{Jup}}$ match the SED of  HD\,95086\,b (Fig.\,\ref{fig: WS}).  For much of this range ($S_{init}=9.5-13$) a mass of $4\,\mathrm{M_{Jup}}$ is favored.

\section{Conclusions}
We reported the near-IR detections of HD\,95086\,b from GPI public commissioning data. We confirmed that the companion is comoving with HD\,95086 and derived the first estimates of its magnitudes with respect to its star: H = $13.1\pm0.9$ and $\mathrm{K}_1$ = $12.1\pm0.5$.

While the mid-IR luminosity of HD\,95086\,b is best consistent with an L-T transition object, it has redder near-IR colors than other young, imaged planet-mass companions.  This is consistent with a very dusty and low surface gravity atmosphere.

Comparison with atmosphere models provide $\mathrm{600\:K \leq T_{\mathrm{eff}} \leq 1500\:K}$ and  $\mathrm{2.1\:dex \leq log\:g \leq 4.5\:dex}$. Evolutionary models are consistent with a mass of $5\pm2$\,M$_{Jup}$.  However, the models are affected by systematic errors that are difficult to quantify because of the lack of young objects at the L/T transition.

More higher precision spectroscopic and photometric data for HD\,95086\,b are required to refine the planet properties. 
\\

{\small Acknowledgments: we thank the consortium who built the GPI instrument and the data analysis team for developing reduction tools. We are grateful to Dave Spiegel and Adam Burrows for making the warm-start models publicly avaliable. JR, MB, GC, and AML acknowledge financial support from the French National Research Agency (ANR) through project grant ANR10-BLANC0504-01. This research has benefitted from the SpeX Prism Spectral Libraries, maintained by Adam Burgasser at http://pono.\-ucsd.edu\-/$\sim$adam/brown\-dwarfs/spex\-prism. JLB PhD is funded by the LabEx “Exploration Spatiale des Environnements Plan\'etaires” (ESEP) \# 2011-LABX-030. TC is supported by a McLean Postdoctoral Fellowship.}


\begin{thebibliography}{...}   

    \bibitem[Allard et al. (2012)]{Allard12}
	Allard, F., Batten, A., Budding, E., et al.\ 2012, {\sl IAU Proceedings}, 282, 235.

    \bibitem[Baraffe et al. (2003)]{Baraffe03}
	Baraffe, I., Chabrier, G., Barman, T., Allard, F., \& Hauschildt, P.\  2003,  {\sl Astronomy and Astrophysics}, 402, 701.

     \bibitem[Barman et al. (2011)]{Barman2011} 
          Barman, T., Macintosh, B., Konopacky, Q. M., Marois, C., 2011, \apj, 735, L39     

    \bibitem[Baudino et al.(2014)]{Baudino14}
      Baudino, J.-L, Br\'ezard, B., Boccaletti, A., Bonnefoy, M., Lagrange, A.-M.\ 2014, {\sl IAU Proceedings}, 299, 277.

    \bibitem[Boccaletti et al.(2012)]{boccaletti12}
      Boccaletti, A., Augereau, J.-C., Lagrange, A.-M., et al.\ 2012, \aap, 544, 85

     \bibitem[Bonnefoy et al. (2013)]{Bonnefoy13}
	Bonnefoy, M., Boccaletti, A., Lagrange, A.-M., et al.\ 2013, \aap,  555, 107
	
     \bibitem[Bonnefoy et al. (2014)]{Bonnefoy14}
          Bonnefoy, M., Chauvin, G., Lagrange, A.-M., et al.\ 2014, \aap, 562, 127
          
      \bibitem[Chabrier et al. (2000)]{Chabrier00}
	Chabrier, G., Baraffe, I., Allard, F., Hauschildt, P.\  2000,  \apj, 542, 464.

      \bibitem[Chauvin et al. (2004)]{Chauvin04}
	Chauvin, G., Lagrange, A.-M., Dumas, C., et al.\ 2004, \aap,  425, 29.

    \bibitem[Chauvin et al.(2012)]{chauvin12}
      Chauvin, G., Lagrange, A.-M., Beust, H., et al.\ 2012, \aap, 542, A41.
      
    \bibitem[Chun et al.(2008)]{chun08}
      Chun, M., Toomey, D., Wahhaj, Z., et al.\ 2008, \procspie, 7015, 70151V.
      
    \bibitem[Currie et al.(2011)]{Currie2011}
          Currie, T., Burrows, A., Itoh, Y., et al., 2011, \apj, 729, 128
          
     \bibitem[Currie et al.(2013)]{currie13}
 	     Currie, T., Burrows, A., Madhusudhan, N., et al.\ 2013, \apj, 776, 15
     \bibitem[Currie et al.(2014)]{currie14}
               Currie, T., Daemgen, S., Debes, J. H., et al., 2014, \apj, 780, L30     
     \bibitem[Cutri et al.(2003)]{cutri03}
 	 Cutri, R., Skrutskie, M., Van Dyk, S., et al.\ 2003, VizieR On-line Data Catalog: II/246. 	

     \bibitem[Cutri et al.(2012)]{cutri12}
 	  Cutri, R., Skrutskie, M., Van Dyk, S., et al.\ 2012, VizieR On-line Data Catalog: II/311.

    \bibitem[Dahn et al.(2002)]{Dahn02}
      Dahn, C., Harris, H., Vrba, F., et al.\ 2002, \apj, 124, 1170.
 
    \bibitem[Galicher et al.(2012)]{galicher12}
      Galicher, R., \& Marois, C.\ 2012, proceedings of the AO4ELT conference, P25.

    \bibitem[Lafreni{\`e}re et al.(2007)]{lafreniere07}
      Lafreni{\`e}re, D., Marois, C., Doyon, R., Nadeau, D., \& Artigau, {\'E}.\ 2007, \apj, 660, 770.

    \bibitem[Lagrange et al.(2010)]{lagrange10}
      Lagrange, A.-M., Bonnefoy, M., Chauvin, G., et al.\ 2010, Science, 329, 57. 
      
    \bibitem[Leggett et al. (2010)]{Leggett10}
	Leggett, S., Burningham, B., Saumon, D., et al.\ 2010,  \apj, 710, 1627.

    \bibitem[Leggett et al. (2013)]{Leggett13}
	Leggett, S., Morley, C., Marley, M., et al.\ 2013, \apj, 763, 130.

    \bibitem[Lenzen et al.(2003)]{lenzen03}
      Lenzen, R., Hartung, M., Brandner, W., et al.\ 2003, \procspie, 4841, 944.

    \bibitem[Marley  et al.(2007)]{marley07}
      Marley, M., Fortney, J., Hubickyj, O., Bodenheimer, P., Lissauer, J.\ 2007, \apj, 655, 541.      

    \bibitem[Marley  et al.(2012)]{Marley12}
     Marley, M., Saumon, D., Cushing, M. et al. \ 2012, \apj, 754, 135.      

     \bibitem[Marois et al.(2006a)]{marois06a}
      Marois, C., Lafreni{\`e}re, D., Doyon, R., Macintosh, B., \& Nadeau, D.\ 2006a, \apj, 641, 556. 
    
    \bibitem[Marois et al.(2006b)]{marois06b}
      Marois, C., Lafreni{\`e}re, D., Macintosh, B.,  \& Doyon, R.\ 2006b, \apj, 647, 612. 

        \bibitem[Marois et al. (2008)]{Marois08}
	Marois, C., Macintosh, B., Barman, T., et al.\ 2008,  {\sl Science},  322, 1348.

         \bibitem[Marois et al.(2010a)]{Marois10a}
	Marois, C., Macintosh, B., V\'eran, J.-P.\ 2010a, \procspie,  7736, 77361.

         \bibitem[Marois et al.(2010b)]{Marois10b}
	Marois, C., Zuckerman, B., Konopacky, Q., et al.\ 2010b,  {\sl Nature},  468, 1080.
  
    \bibitem[Marois et al.(2014)]{marois14}
      Marois, C., Correia, C., V\'{e}ran, J.-P., Currie, T.\ 2014, {\sl IAU symposium}, 299, 48-49. 

    \bibitem[Macintosh et al.(2014)]{macintosh14}
      Macintosh, B., Graham, J., Ingraham, P. et al.\ 2014, {\sl accepted in the Proceedings of the National Academy of Sciences of the United States of America}.

    \bibitem[Meshkat et al.(2013)]{meshkat13}
      Meshkat, T., Bailey, V., Rameau, J., et al.\ 2013, \apjl, 775, L40.

    \bibitem[Mordasini et al.(2012)]{mordasini12}
      Mordasini, C., Alibert, Y., Georgy, C., et al.\ 2012, \aa, 547, A112.

    \bibitem[Pecaut et al.(2012)]{pecaut12}
      Pecaut, M.-J., Mamajek, E. E., \&  Bubar, E. J.\ 2012, \apj, 746, 154.

    \bibitem[Perrin et al.(2014)]{perrin14}
      Perrin, M. \& GPI Instrument and Science teams\ 2014, American Astronomical Society Meeting Abstracts, 223, 348.

    \bibitem[Pickles et al.(1998)]{pickles98}
      Pickles \ 1998, PASP, 110, 863.

    \bibitem[Racine et al.(1999)]{racine99}
      Racine, R., \& Nadeau, D. \& Doyon, R. \ 1999, ESO conference Proceedings, 56, 377.

   \bibitem[Rameau et al.(2013a)]{rameau13a}
     Rameau, J., Chauvin, G., Lagrange, A.-M., et al.\ 2013a, \aap, 772, L15.

    \bibitem[Rameau et al.(2013b)]{rameau13b}
      Rameau, J., Chauvin, G., Lagrange, A.-M., et al.\ 2013b, \aap, 779, L26.

    \bibitem[Rousset et al.(2003)]{rousset03}
      Rousset, G., Lacombe, F., Puget, P., et al.\ 2003, \procspie, 4839, 140--149.

    \bibitem[Soummer et al.(2012)]{soummer12}
      Soummer, R., Pueyo, L., \& Larkin, J.\ 2012, \apjl, 755, L28.

    \bibitem[Spiegel\&Burrows(2012)]{SB12}
      Spiegel, D. \& Burrows, A.\ 2012, \apj, 745, 174 

    \bibitem[Stephens et al.(2009)]{Stephens09}
      Stephens, D., Leggett, S., Cushing, M., Marley, M., Saumon, D., Geballe, T., Golimowski, D., Fan, X., Noll, K.\ 2009, \apj, 702, 154 

    \bibitem[van Leeuwen et al.(2007)]{vanLeeuwen07}
      van Leeuwen, F. 2007,  \aap, 474, 653.

    \bibitem[Zurlo et al.(2013)]{zurlo13}
      Zurlo, A., Vigan, A., Hagelberg, J., et al.\, 2013, \aa, 554., A21.
  
   \end{thebibliography}
\end{document}